 \documentclass[10pt, conference, letterpaper]{IEEEtran}
\usepackage{amssymb}
\usepackage{cite}

\ifCLASSINFOpdf
  \usepackage[pdftex]{graphicx}
  \graphicspath{{../pdf/}{../jpeg/}}
  \DeclareGraphicsExtensions{.pdf,.jpeg,.png}
\else
   \usepackage[dvips]{graphicx}
   \graphicspath{{../eps/}}
   \DeclareGraphicsExtensions{.eps}
\fi
\usepackage{amsmath}
\usepackage{array}

\ifCLASSOPTIONcompsoc
  \usepackage[caption=false,font=normalsize,labelfont=sf,textfont=sf]{subfig}
\else
  \usepackage[caption=false,font=footnotesize]{subfig}
\fi
\usepackage{microtype}

\begin{document}
%
\title{Pull-based Bloom Filter-based Routing for Information-Centric Networks}

\author{\IEEEauthorblockN{Ali~Marandi\IEEEauthorrefmark{1}, Torsten~Braun\IEEEauthorrefmark{1}, Kav\'e Salamatian\IEEEauthorrefmark{2} and Nikolaos~Thomos\IEEEauthorrefmark{3} 
}
\IEEEauthorblockA {
\IEEEauthorrefmark{1}University of Bern, Bern, Switzerland \\ Email:\{marandi, braun@inf.unibe.ch\}\\
 \IEEEauthorrefmark{2}Universit\'e de Savoie, France\\Email: kave.salamatian@univ-savoie.fr\\
 \IEEEauthorrefmark{3}University of Essex, Colchester, United Kingdom\\ Email: nthomos@essex.ac.uk\\
}}


\maketitle


\begin{abstract}
In Named Data Networking (NDN), there is a need for routing protocols to populate Forwarding Information Base (FIB) tables so that the Interest messages can be forwarded. To populate FIBs, clients and routers require some routing information. One method to obtain this information is that network nodes exchange routing information by each node advertising the available content objects. Bloom Filter-based Routing approaches like BFR \cite{bfr}, use Bloom Filters (BFs) to advertise all provided content objects, which consumes valuable bandwidth and storage resources. This strategy is inefficient as clients request only a small number of the provided content objects and they do not need the content advertisement information for all provided content objects. In this paper, we propose a novel routing algorithm for NDN called {\emph{pull-based BFR}} in which servers only advertise the demanded file names. We compare the performance of pull-based BFR with original BFR and with a flooding-assisted routing protocol. Our experimental evaluations show that pull-based BFR outperforms original BFR in terms of communication overhead needed for content advertisements, average roundtrip delay,  memory resources needed for storing content advertisements at clients and routers, and the impact of false positive reports on routing. The comparisons also show that pull-based BFR outperforms flooding-assisted routing in terms of average round-trip delay.
\end{abstract}


%
\IEEEpeerreviewmaketitle

\section{Introduction}
\label{sec::intro}
In this paper, we consider NDN \cite{ndnref} as one of the most prominent Information-Centric Networking (ICN)\cite{ICNsurvey} architectures. In NDN, FIB population is the prerequisite phase for routing Interest messages. At network setup, i.e., at the begining of network operation, the FIBs are empty. The adopted routing protocol is responsible for FIB population. One method of FIB population is to use content advertisements. Such a technique is BFR \cite{bfr}, which is a routing protocol for NDN that uses BF-based content advertisements for FIB population. In BFR\cite{bfr}, servers compactly represent the provided file names using Bloom Filters (BFs) and push them to the network using Content Advertisement (CA) messages. Clients and routers store the content advertisement BFs to populate the FIBs so that they can route the Interest messages they receive. However, by increasing the size of content universe (the number of available content objects in the network), there is a linear growth in the communication overhead required for propagating all the CA messages, and the memory space that clients and routers need to store all the CA messages. 

To overcome BFR shortcomings, in this paper,  we propose a new routing protocol called {\emph{pull-based Bloom Filter-based Routing}}. In pull-based BFR, servers only advertise the demanded file names. Thus, pull-based BFR needs significantly less bandwidth for content advertisements than push-based BFR (original BFR\cite{bfr}). Further, in pull-based BFR, clients and routers do not need to store the content advertisements for the entire content universe in contrast to push-based BFR. Therefore, clients and routers need significantly less memory resources to store content advertisements using pull-based BFR than push-based BFR. Pull-based BFR is a fully-distributed, content oriented, and topology oblivious protocol, like push-based BFR.

We compare the performance of pull-based and push-based BFR protocols. The comparison shows that pull-based BFR outperforms push-based BFR in terms of content advertisement overhead, average round-trip delay, memory resources needed for storing content advertisements at clients and routers, and the impact of false positive reports of BFs on routing. For the sake of completeness, we also compare the performance of the proposed pull-based BFR with a Flooding-assisted Routing (FaR) protocol. The results make clear that pull-based BFR outperforms FaR in terms of average round-trip delay. 

The rest of this paper is structured as follows. We give an overview to push-based BFR in Section \ref{sec::bfr}. Next, in Section \ref{sec::pull-based}, we describe pull-based BFR. Section \ref{sec::perf} discusses performance evaluation. Then, we briefly discuss related works in Section \ref{sec::rw}. Section \ref{concl} concludes the paper.

\section{Push-based Bloom Filter-based Routing}
\label{sec::bfr}
Bloom Filters (BFs) are used to compactly represent sets and find applications in IP networking, e.g., finding Longest Prefix Match, probabilistic routing algorithms, summary cache exchange, and matching IP addresses \cite{bfappssurvey}. BFs have been used in NDN for similar purposes with IP systems \cite{bfr,iscc,mapit,cachsharing,ifibf}. A BF is a bit vector initialized to zero. The bit values of a BF are set with the help of a number of hash functions. In particular, to insert an element into a BF, one has to give the element as an input to the hash functions. The output values of the hash functions specify the indices of the bits to be set in the bit vector. To check whether a BF contains an element, all the bits specified by the hashes of the element have to be set. When BFs are used, it is impossible to have a false negative report, i.e., to falsely report that the BF does not contain an element, but a false positive report may happen, i.e., to falsely report that the BF contains an element, with the probability $p$. If $n$ is the inserted element count of the BF, $m$ is the bit vector's size, and $k$ is the number of hash functions, the trade-offs between $m$, $k$, $p$, and $n$ are given as in \cite{bloom}:
\begin{equation}
\label{MvaK}
m=-\frac{nln(p)}{(ln2)^2}, \\
k=\frac{m}{n}ln2
\end{equation}
There are two advantages of using a BF for representing a set than using a regular array: 1) compressed representation of the set, and 2) less complex element search. The complexity of searching an element in a BF is $O(k)$, whereas the complexity of searching an element in a regular array is $O(n)$ with $k<<n$.

\begin{figure}[t]
  \vspace{-1em}
\centering
 \includegraphics[width=0.75\columnwidth]{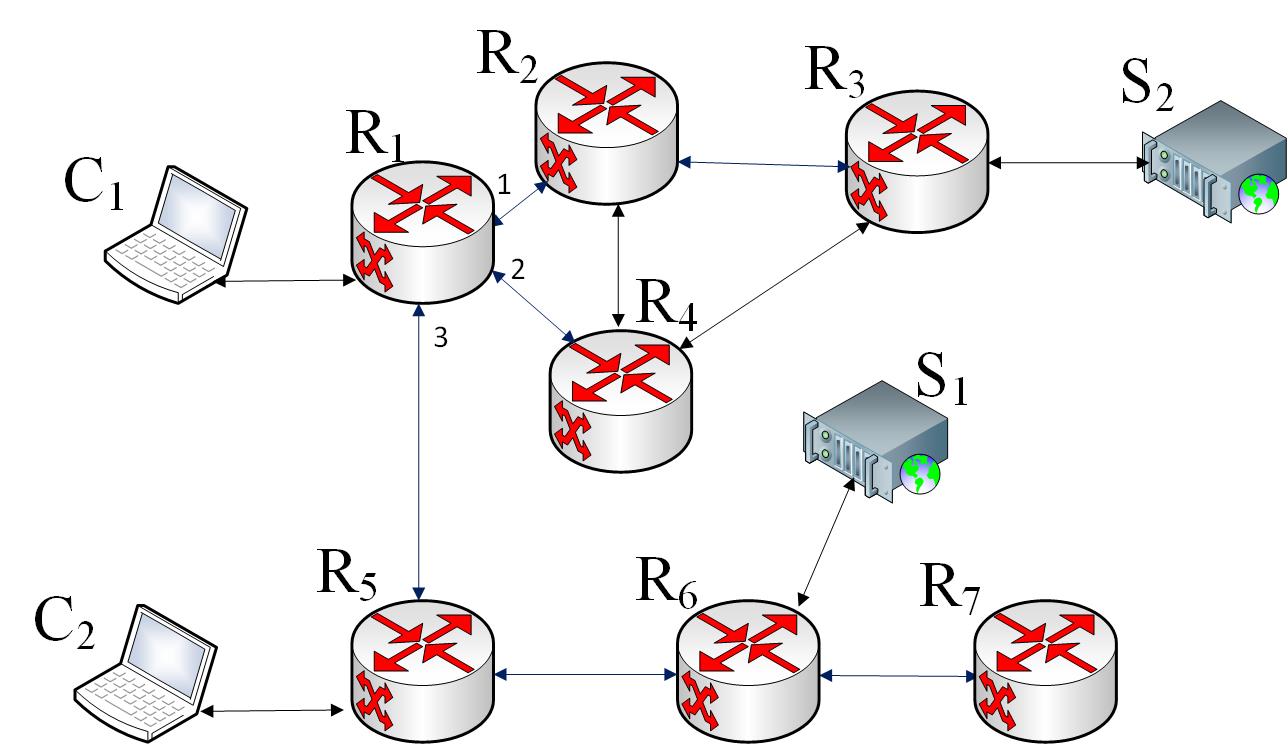}
 \caption{A topology for describing push-based BFR.}
 \label{fig::CAp}
\end{figure}

\begin{figure}[!t]
    \centering
        \subfloat[A PIT entry stored at router $R_1$]{\label{fig::in-rec}\includegraphics[width=0.7\columnwidth]{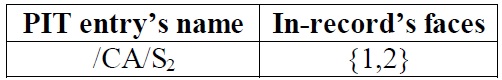}}
        \\
       \subfloat[A FIB entry stored at router $R_1$ for name prefix $N_1$ ]{\label{fig::fib}\includegraphics[width=0.7\columnwidth]{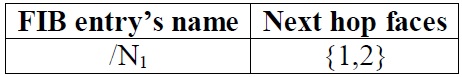}}
    \label{fig:subfigname}
    \caption{Structure of PIT and FIB entries.}
\vspace{-1.5em}
\end{figure}

Push-based BFR \cite{bfr} operates in two phases: 1) representation and advertisement of content objects using BFs, 2) FIB population and content retrieval. Let us describe the first phase using Fig. \ref{fig::CAp}. In this figure, when server $S_2$ produces some files, it then inserts the produced file names as well as their available name prefixes into a BF. To advertise these file names, server $S_2$ needs to broadcast a message that encapsulates the resulting BF. For this purpose, push-based BFR uses a new Interest message type called CA message, which has the name prefix $/CA/serverID$. CA messages contain the required information to retrieve BFs, a nonce value, and a lifetime.
Note that servers do not broadcast CA messages to demand any content objects, but they only broadcast these messages to inform other nodes about the content objects they possess. Similar to Interest messages, NDN Forwarding Daemon (NFD) is used for loop detection as well as for duplicate detection and discard of CA messages. When a router or a client receives a CA message, it stores this message in the PIT and updates the ID of the face over which the CA message has been received in the in-record field of the PIT entry. For example in Fig. \ref{fig::CAp}, router $R_1$ receives the CA message of server $S_2$ over faces $1$ and $2$. The corresponding structure of the PIT entry is presented in Fig. \ref{fig::in-rec}, where the in-record for name $/CA/S_2$ is stored at router $R_1$. The information stored in the in-record field is useful for FIB population later. 
To describe FIB population and routing processes, in Fig. \ref{fig::CAp} we assume that client $C_1$ receives and stores the CA message of server $S_2$ at time instant $t_1$. Further, we assume that client $C_1$ issues Interest $I_1$ to demand a segment of file name $N_1$, after time instant $t_1$. To populate the FIB for name $N_1$ and to route Interest $I_1$, client $C_1$ retrieves the BF of the CA message of server $S_2$ from the BF information stored in the CA message and checks name $N_1$ against the BF. If name $N_1$ exists in the BF of server $S_2$'s CA message, client $C_1$ checks the in-record information to know the face(s) over which server $S_2$'s CA message has been received. In Fig. \ref{fig::CAp}, for client $C_1$, there is only one face, thus it forwards Interest $I_1$ towards router $R_1$, but as Fig. \ref{fig::in-rec} shows, router $R_1$ has received server $S_2$'s CA message over two faces, i.e., faces $1$ and $2$. Thus, this router considers both faces $1$ and $2$ as the next hop faces for name prefix $N_1$ and populates the FIB for this name prefix as Fig. \ref{fig::fib} shows. Push-based BFR uses the multicast forwarding strategy, because it is a multi-path routing protocol.

Note that core routers' processors have several cores. Thus, a core router can chack names against multiple BFs in parallel, thus this does not create any performance issues.
%
\section{Pull-based Bloom Filter-based Routing}
\label{sec::pull-based}
The rationale behind designing a pull-based BFR method is to advertise only the demanded content objects. 
When servers only advertise the demanded content objects, it is expected that: 1) significant amount of bandwidth will be saved, and 2) other network nodes (clients and routers) will need significantly less memory space to store content advertisement information. The adoptation of this content advertisement strategy can resolve scalability issues of push-based BFR, as in push-based BFR servers advertise all the file names of the content universe. The main difficulty arising from advertising only the demanded content objects is that servers do not know a priori which content objects will be demanded. To overcome this problem, in pull-based BFR, we follow a BF-based strategy to inform the servers about the demanded file names, which we will explain in the next sub-section.
\subsection{Pull-based BFR's Operation}
Content advertisement in pull-based BFR is performed in two consecutive phases: 1) clients and routers use a BF-based strategy to inform the servers about the demanded file names, and 2) servers proceed with the advertisement of these names using CA messages. Upon reception of CA messages, clients and routers store the content advertisement information and populate the FIBs for pending Interests to route them. To summarize, pull-based BFR's operation is done in three stages: 1) pulling content advertisements, 2) content advertisement, and 3) FIB population and content retrieval. 

\begin{figure}[t]
  \vspace{-0.5em}
\centering
 \includegraphics[width=1\columnwidth]{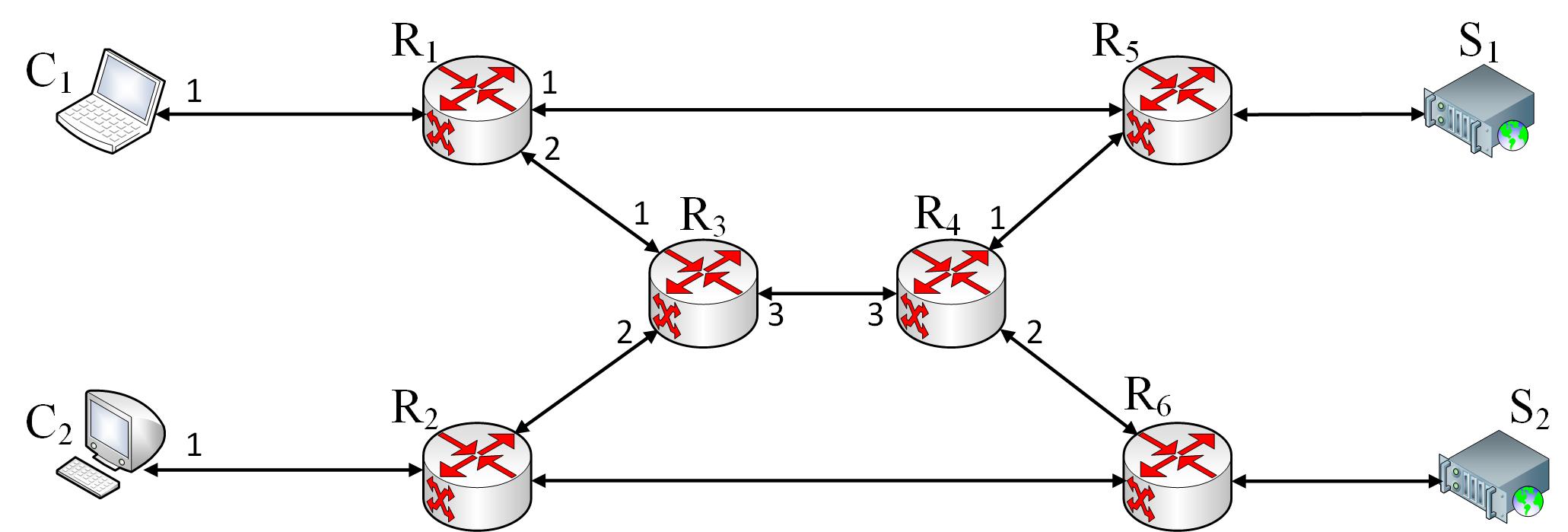}
 \caption{A topology for describing pull-based BFR.}
 \label{fig::ibf}
 \vspace{-1em}
\end{figure}

Let us explain our BF-based method of informing servers about the demanded file names with the help of Fig. \ref{fig::ibf}. In Fig. \ref{fig::ibf}, we assume that client $C_1$ issues Interest $I_1$ to retrieve a segment of file name $N_1$ under the following conditions: 1) there is no FIB entry for $N_1$ or a name prefix of it, and 2) there is no stored content advertisement BF that contains $N_1$ or a name prefix of it. Thus, client $C_1$ avoids forwarding Interest $I_1$ and keeps it as pending. Nevertheless, client $C_1$ informs the servers that file name $N_1$ is demanded to pull the content advertisement information for it. For this purpose, client $C_1$ creates a BF, which contains file name $N_1$ as well as all its name prefixes and creates a Content Advertisement Request (CAR) message of type Interest called $CAR_{C_1}$ with name $/CAR/C_1/sequenceNumber$ that encapsulates the BF. Then, client $C_1$ broadcasts  $CAR_{C_1}$ to inform the servers about the demanded file names and to pull the needed content advertisements. When a router receives a CAR message, it waits for an {\em{aggregation threshold time}} $\delta$ to receive other CAR messages issued by other clients. Assume that client $C_2$ issues Interest $I_2$ to demand a segment of file name $N_2$ for which no FIB entry and no content advertisement information is available. Thus, client $C_2$ broadcasts a CAR message called $CAR_{C_2}$ with name $/CAR/C_2/sequenceNumber$ carrying a BF that contains file name $N_2$ as well as all its name prefixes. If router $R_3$ receives the CAR messages of clients $C_1$ and $C_2$, within a time interval $\delta$, it forwards $CAR_{C_1}$  and $CAR_{C_2}$ over faces $1$ and $2$, respectively. At the same time, router $R_3$ forwards the aggregation of $CAR_{C_1}$  and $CAR_{C_2}$ over face $3$. To aggregate $CAR_{C_1}$ and $CAR_{C_2}$, router $R_3$ makes a union of their BFs and puts the resulting BF into a new CAR message with name $/CAR/aggregated/R_3/sequenceNumber$. Then router $R_3$ forwards this message over face $3$. Therefore, as a rule of thumb, a router forwards over face $f$ the union of BF(s) that have been received over other faces and have not been sent over face $f$ before. When router $R_3$ forwards message $/CAR/aggregated/R_3/sequenceNumber$ over face $3$, router $R_3$ updates the out-records of both messages $CAR_{C_1}$ and $CAR_{C_2}$ by adding face $3$ to record that both these messages have been forwarded over face $3$. Further, router $R_3$ will not use message $/CAR/aggregated/R_3/sequenceNumber$ in future aggregations, because the third name component specifies that this message is created by router $R_3$ itself and has not been received from other nodes. Routers $R_4$, $R_5$, and $R_6$ follow the same forwarding process for CAR messages. Nodes make use of a {\emph{sequence number counter}} for calculating the sequence numbers of CAR messages. 

To permit BF union operations, we assume that all nodes create the BFs of the CAR messages with the same size, and that they generate the hash functions using a universal seed, i.e., all nodes use the same set of hash functions for BFs. 
In (\ref{MvaK}), if we assign a constant value to $m$ and we specify the value of $p$, we will derive the maximum optimal value for $n$, which estimates the maximum number of requested file names that can be inserted into the BF. It is not a problem that all nodes use a universal seed to generate the hash functions for the BFs of all CAR messages, as all nodes can use a well-known word, e.g., {\em{NDN}} as the universal seed to generate hash functions. 
When servers $S_1$ and $S_2$ receive a CAR message, they check all the produced file names against the BF of the received CAR message\footnote{We assume that servers have multi-core processors and can check multiple names against multiple BFs in parallel. Thus, this does not create a performance issue.}. The file names that exist in the BF of the CAR message are the demanded file names that should be advertised. Thus, both servers $S_1$ and $S_2$ first create a list of these file names called {\em{toBeAdvertisedList}} and then a BF called {\em{toBeAdvertisedBF}} with size equal to that of the received CAR message's BF. When a server notes that a produced file name exists in the BF of the CAR message, it inserts the file name into the BF  {\em{toBeAdvertisedBF}}. Then, the server creates a CA message, from type Interest with name prefix $/CA/serverID/sequenceNumber$ carrying the {\em{toBeAdvertisedBF}}. The server broadcasts the CA message to the network to advertise the demanded content object and not to demand any content objects. In our example, if router $R_4$ receives the CA messages of servers $S_1$ and $S_2$, namely, $CA_{S_1}$ and $CA_{S_2}$, which have the names $/CA/S_1/sequenceNumber$ and $/CA/S_2/sequenceNumber$, respectively, within a time interval $\delta$, Router $R_4$ forwards $CA_{S_1}$ and $CA_{S_2}$ over faces $1$ and $2$, respectively. Router $R_4$ aggregates $CA_{S_1}$ and $CA_{S_2}$ unioning their BFs and places the resulting BF into an aggregated message, which has the name $/CA/aggregated/R_4/sequenceNumber$ and forwards this message over face $3$. 

When clients $C_1$ and $C_2$ receive the CA message, they can populate their FIBs for name prefixes $N_1$ and $N_2$, which allows them to route Interests $I_1$ and $I_2$. When routers receive Interests  $I_1$ and $I_2$ from the clients, they also populate the FIBs using the stored CA messages and continue routing the Interests until the demanded content objects are retrieved.

\subsection{Bloom Filter Aggregation}
\label{subsec::aggrLimit}
If a router makes a union of the BFs $BF_1$ and $BF_2$, which are not subset or equal to each other, i.e., $(BF_1 \nsubseteq BF_2) \land (BF_2\nsubseteq BF_1)$, the number of $1$s in the bit vector of the resulting BF $BF_{union}$ will be greater than the number of $1$s in each of $BF_1$ and $BF_2$. Thus, if routers do not stop unioning BFs that are not subset or equal to each other, at some point all the bits of the bit vector of the resulting BF will be set to $1$. Such a BF does not function properly because it falsely claims that it contains all the existing names. Therefore, routers should stop unioning the BFs of both CAR and CA messages according to the maximum capacity of BFs. As we explained before, we consider a constant size $m$ as well as a probability of false positive error $p$ for the BFs of CAR messages. Then, using ($\ref{MvaK}$), we calculate $n$, which is the maximum capacity of the BF.
Assume that router $R$ wants to aggregate $BF_1$ and $BF_2$, which have inserted element counts $|BF_1|$ and $|BF_2|$, respectively. First, router $R$ checks if $BF_1$ and $BF_2$ are identical. For this purpose, router $R$ makes an $XOR$ of the bit vectors of $BF_1$ and $BF_2$. If all the bits of the resulting bit vector are zero, $BF_1$ and $BF_2$ are identical.  In such a case, there is no need to make a union of them. The second check is to examine whether the following proposition is true $(BF_1 \subset BF_2) \lor (BF_2 \subset BF_1)$. For this purpose, router $R$ calculates $BF_{intersection} = BF_1 \cap BF_2$. If the resulting bit vector is identical with the bit vector of $BF_1$, it means that $BF_1 \cup BF_2 = BF_2$. In this case, again router $R$ does not need to calculate the union of $BF_1$ and $BF_2$. If $(BF_1 \nsubseteq BF_2) \land (BF_2\nsubseteq BF_1)$, then router $R$ makes a union of $BF_1$ and $BF_2$. In this case, if $BF_{union} = BF_1 \cup BF_2$ and $BF_{intersection} = BF_1 \cap BF_2$, theoretically we have $|BF_{union}| = |BF_1|+| BF_2|-|BF_{intersection}|$. However, practically it is not possible to calculate $|BF_{intersection}|$, precisely. Therefore, router $R$ sets $|BF_{union}|=|BF_1|+|BF_2|$, which is a conservative upper bound. If $|BF_1|+|BF_2| < n$, router $R$ will aggregate $BF_1$ and $BF_2$. Otherwise, router $R$ avoids aggregating these BFs.
\subsection{The Impact of False Positive Errors on Pull-based BFR's Operation}
\label{subsec::fp}
The impact of false positive errors on the operation of pull-based BFR should be considered in two cases: 1) if servers check the produced file names against the CAR messages BFs, 2) if clients or routers check the pending Interest names against the CA messages BFs. Consider in Fig. \ref{fig::ibf} that server $S_1$ receives a CAR message carrying a BF, which contains names $N_1$ and $N_2$. If server $S_1$ checks the file name $N_3$ against the received BF and the BF gives a false positive report, server $S_1$ will insert name $N_3$ into the BF of the CA message $/CA/S_1$ and advertises this message. Therefore, the CA message $/CA/S_1$ advertises the file name $N_3$, which has not been demanded. This is not a problem because it is guaranteed that no false negative errors happen using BFs, and, therefore, servers advertise the produced file names that are demanded anyways. Let us again examine Fig. \ref{fig::ibf} to discuss the impact of false positive reports from the BFs of CA messages, when clients or routers check the Interest names against these BFs for FIB population and routing purposes. In Fig.  \ref{fig::ibf}, we assume that router $R_4$ checks the name $N_i$ for Interest $i$ against the BF of CA message $CA_{S_1}$ issued by server $S_1$
. If the BF gives a false positive report, router $R_4$ will forward the Interest $i$ over face $1$. Consequently, Interest $i$ will be routed towards a wrong server, i.e., server $S_1$. 
When server $S_1$ receives Interest $i$, it sends back a ``No Data'' Nack message \cite{stateful} to inform router $R_4$ that server $S_1$ does not store the Data that Interest $i$ requests. When router $R_4$ receives the Nack message, it will remove from the FIB the incorrect next hop information corresponding to name $N_i$. Further, if Interest $i$ is not satisfied yet, router $R_4$ will send a CAR message containing $N_i$ to receive the correct routing information.

%
%
\begin{figure*}[!t]
 \centering
 \subfloat[][]{\label{fig::delta}\includegraphics[width=0.33\textwidth]{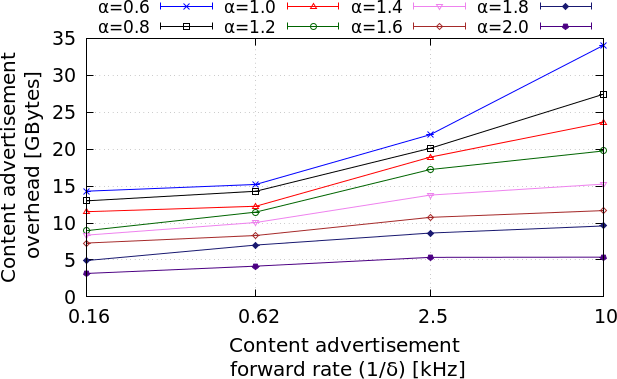}}
 \subfloat[][]{\label{fig::advert-overhead}\includegraphics[width=0.33\textwidth]{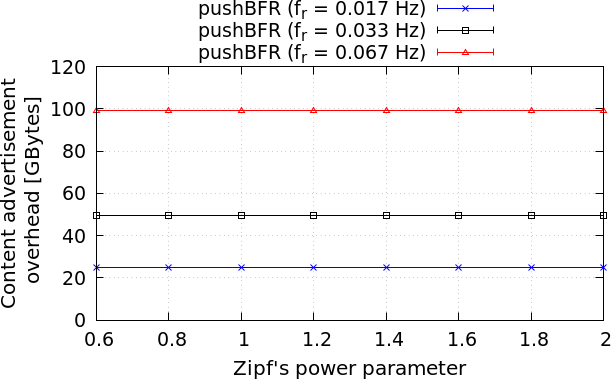}}
 \subfloat[][]{\label{fig::delay-real}\includegraphics[width=0.33\textwidth]{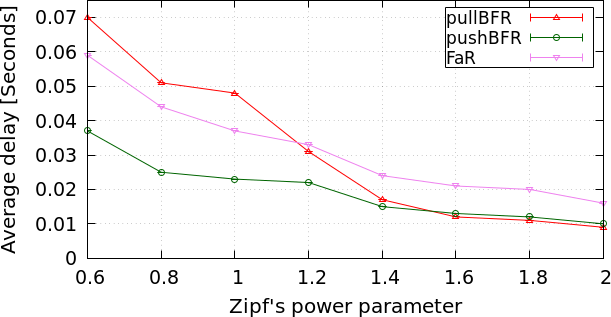}}
 \caption[]{Performance evaluation for different values of $\alpha$ : (a) routing communication overhead for different values of $\delta$ and $\alpha$; (b) content advertisement overhead; (c) average round-trip delay when links have maximum capacities.} 
 \label{fig:res}
 \vspace{-1.5em}
\end{figure*}
\begin{figure*}[!t]
 \centering
 \subfloat[][]{\label{fig::delay-reduced}\includegraphics[width=0.33\textwidth]{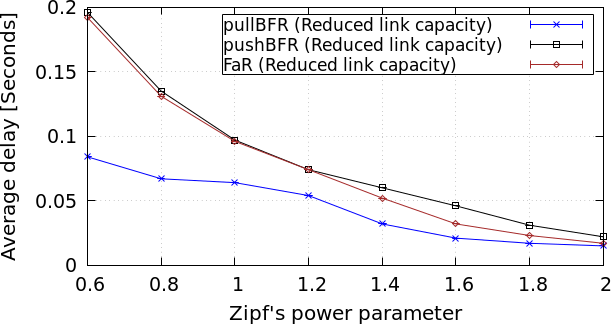}}
 \subfloat[][]{\label{fig::mem}\includegraphics[width=0.33\textwidth]{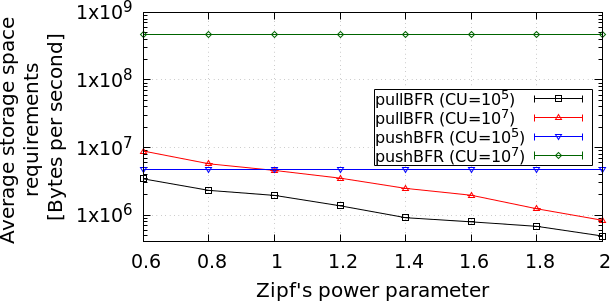}}
 \subfloat[][]{\label{fig::mem2}\includegraphics[width=0.33\textwidth]{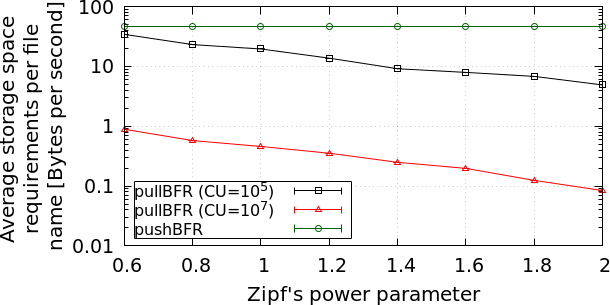}}
 \caption[]{Performance evaluation for : (a) average round-trip delay when links have $20\%$ of their maximum capacities; (b) storage space requirements for storing routing information for different values of $\alpha$.} 
 \label{fig:res2}
 \vspace{-1.5em}
\end{figure*}

\section{Performance Evaluation}
\label{sec::perf}
In this section, we compare the peformance of pull-based BFR, push-based BFR, and flooding-assisted routing protocols implemented in ndnSIM\cite{ndnsim2.1}. 
\subsection{Flooding-assisted Routing}
\label{subsec::flood}
Flooding-assisted Routing (FaR) is a protocol that does not use content advertisements for routing. Thus, clients and routers are not aware of the content objects that each server provides as well as the route to reach them. Hence, at the first phase of routing, clients and routers flood the Interests. Rightafter, when clients and routers receive Data packets, they populate the FIBs for the name prefix of the Data packet. Let us explain flooding-assisted routing with the help of Fig. \ref{fig::CAp}. In Fig. \ref{fig::CAp}, we assume that client $C_1$ issues an Interest $I_1$ to demand a content object with name $N_1$ provided by  server $S_1$. However, client $C_1$ does not have any information about the provider of content object $N_1$. Thus, client $C_1$ floods Interest $I_1$ and waits for the Data packet with name $N_1$. Since server $S_1$ provides the Data packet with name $N_1$,  Router $R_1$ only receives this Data packet over face $3$. Therefore, router $R_1$ populates the FIB for name $N_1$ and considers face $3$ as the only next hop face for name $N_1$ in the FIB.
\subsection{Simulation Settings}
To compare the performance of the protocols under comparison, we use the GEANT topology \cite{geant,bfr}. The topology is built by randomly placing $10$ servers and $50$ clients in the GEANT topology, which connects $40$ routers. Thus, the resulting topology consists $100$ nodes. We assume that the content popularity follows Zipf-Mandelbrot distribution. Equation (\ref{eq::zipf}) shows the probability distribution function for Zipf-Mandelbrot distribution, where $M$ is the cardinality of the content universe and $\alpha$ is the Zipf's power parameter.
 We consider the values of $\alpha$ in the $[0.6,2]$ interval. We use a URL dataset extracted from real HTTP request traces \cite{selfsimilarity}. We assume that the content universe has $100,000$ file names, and that each is divided into $100$ segments. Therefore, there are $10^7$ unique segments. For the BFs of CAR and CA messages, we set $m=716$ Bytes and $P_{fpp}=0.0638$. Recall, that $m$ is the BF's bit vector size and $P_{fpp}$ represents the false positive probability. Hence, using (\ref{MvaK}), the maximum value of $n$ will be $1000$.
\begin{equation}
\label{eq::zipf}
P(x=i) =  \frac{1/i^\alpha}{\sum_{j=1}^{M} 1/j^\alpha}
\end{equation}
\subsection{Performance}
We consider three metrics for assessing the performance of all protocols: 1) content advertisement overhead, 2) average round-trip delay, and 3) storage space requirements for storing routing information. We also compare pull-based and push-based BFR in terms of the impact of false positive reports of BFs on routing.

\subsubsection{Content Advertisement Overhead} 
For pull-based BFR, Fig. \ref{fig::delta} shows the content advertisement overhead, i.e., the total communication overhead required for forwarding CAR and CA messages in terms of {\em{forwarding rate of routing messages}}, which is defined as $\frac{1}{\delta}$. Higher {\em{forwarding rate of routing messages}} results in more frequent forwarding of CAR and CA messages, i.e., less aggregation of CAR and CA messages. We set the $\delta$ values in the $[0.1,6.4]$ interval measured in milliseconds. This results in forwarding rate of routing messages in the $[0.16,10]$ interval in terms of kilohertz. From Fig. \ref{fig::delta}, we observe that for pull-based BFR, the content advertisement overhead increases by increasing the forwarding rate of routing messages, for all $\alpha$ values. 

For push-based BFR, the total communication overhead needed for content advertisements depends on the content universe size, because servers advertise all the file names they produce. However, in pull-based BFR, servers do not advertise the file names that are not demanded. The number of popular files is controlled by the value of $\alpha$ (higher $\alpha$ means less content objects are requested). We observe from Fig. \ref{fig::delta} that for pull-based BFR, the communication overhead needed for content advertisements significantly decreases with higher $\alpha$ values. This is due to the fact that when the value of $\alpha$ increases, less content objects are popular and thus are demanded. Therefore, clients propagate smaller number of CAR messages, because they require less CA information. For push-based BFR, Fig. \ref{fig::advert-overhead} shows the required communication overhead for propagating content advertisements in terms of {\em{content advertisement refresh rate ($f_r$)}}, i.e., the frequency that servers refresh CA messages. From Fig. \ref{fig::advert-overhead}, we observe that for push-based BFR, the communication overhead required for propagating content advertisements increases by increasing $f_r$. When we compare Figs. \ref{fig::delta} and \ref{fig::advert-overhead}, we observe that pull-based BFR requires significantly less communication overhead for propagating content advertisements compared to push-based BFR. For example, in Fig. \ref{fig::advert-overhead}, push-based BFR requires the least communication overhead for propagating content advertisements if $f_r = 0.017 Hz$, however, even in this case,  push-based BFR requires significantly more communication overhead for propagating content advertisements compared to pull-based BFR except when $\frac{1}{\delta}=10 kHz$ and $\alpha$ is in the $[0.6,0.8]$ interval. Note that when pull-based BFR is used and $\frac{1}{\delta}=10 kHz$, nodes perform very little aggregation, which is not of our interest. Therefore, in the rest of results, for pull-based BFR, we use $\frac{1}{\delta}=2.5 kHz$ and for push-based BFR, we use $f_r=0.017 Hz$. For push-based BFR, we use $\frac{1}{f_r}$ as the lifetime of CA messages. For pull-based BFR, we set the lifetime of CAR and CA messages to $4 secs$ and $10 secs$, respectively.
\subsubsection{Average Round-trip Delay} 
From Figs. \ref{fig::delay-real} and \ref{fig::delay-reduced}, we present the results in terms of average round-trip delay, i.e., the average delay that a client experiences from the time it issues an Interest to the time it retrieves the demanded Data packet. We measure this delay for all the studied protocols in two scenarios: 1) when links have full capacity, and 2) when links have only $20\%$ of the original capacity. When users can make use of the full network capacity, Fig. \ref{fig::delay-real} shows that if $\alpha$ is in the  $[0.6, 1]$ interval, push-based BFR performs slightly better than pull-based BFR, because the cardinality of the set of popular content objects is bigger for smaller values of $\alpha$. Thus, pulling content advertisement and CAR aggregation at routers have more impact on the average round-trip delay for pull-based BFR. Nevertheless, when $\alpha$ is in the $[1.2, 2]$ interval, pull-based BFR and push-based BFR perform very close to each other, because much less content objects are popular. Thus, clients need to pull much less CA information and each demanded content object will be cached close to the client that demanded it after its first retrieval. Therefore, in this case, the delay caused by pulling content advertisements and aggregating CARs and CAs has less impact on the overall average round-trip delay. When the links of the GEANT topology \cite{geant} have $100\%$ of their capacities and $\alpha$ is in the $[0.6,1]$ interval, FaR performs close to pull-based BFR. However, when $\alpha$ is in the $[1.2,2]$ interval, pull-based BFR outperforms FaR. We observe from Fig.  \ref{fig::delay-reduced} that by reducing link capacity, push-based BFR and FaR protocols are more affected, while we observe the smallest impact of limited link capacity on the performance of pull-based BFR. The reason is that pull-based BFR aggregates CAR and CA messages and, therefore, it has much less number of transmissions than push-based BFR and FaR.
\subsubsection{Storage Space Requirements for Storing Routing Information}
\label{subsec:mem}
Routing information for push-based BFR consists of CA messages, while for pull-based BFR, routing information includes both CA and CAR messages. Fig. \ref{fig::mem} compares pull-based and push-based BFR in terms of average storage space a node requires to store routing information per second. For push-based BFR, we observe from Fig. \ref{fig::mem} that the storage space requirements for storing routing information significantly increases with the size of Content Universe (CU). As explained before, the reason is that using push-based BFR, clients and routers require to store the routing information for the entire CU. However, using pull-based BFR, the nodes only store the routing information for the demanded file names. Therefore, from Fig. \ref{fig::mem} we observe that the storage space requirements for pull-based BFR slightly grows when we increase CU size from $10^5$ to $10^7$. Fig.  \ref{fig::mem} also shows that the storage space requirements for pull-based BFR is controlled by the value of $\alpha$, meaning that for higher $\alpha$ values, pull-based BFR has less storage space requirements, while the storage space requirements for push-based BFR only depends on the CU size. Fig. \ref{fig::mem} shows that for both $CU=10^5$ and $CU=10^7$, pull-based BFR requires significantly less storage space for storing routing information compared to push-based BFR. Nevertheless, we observe from Fig. \ref {fig::mem} that when the CU size grows from $10^5$ to $10^7$, pull-based BFR outperforms push-based BFR more significantly. Fig. \ref{fig::mem2} compares pull-based and push-based BFR in terms of the average storage space that a node needs to store routing information for one file name per second. We observe from Fig. \ref{fig::mem2} that pull-based BFR outperforms push-based BFR for both $CU=10^5$ and $CU=10^7$. If the CU size grows from $10^5$ to $10^7$, Fig. \ref{fig::mem2} shows that pull-based BFR outperforms push-based BFR more significantly.
\subsubsection{Impact of False Positive Reports on Routing}
\label{subsec:fpImpact}
\begin{figure}[t]
\centering
 \includegraphics[width=0.8\columnwidth]{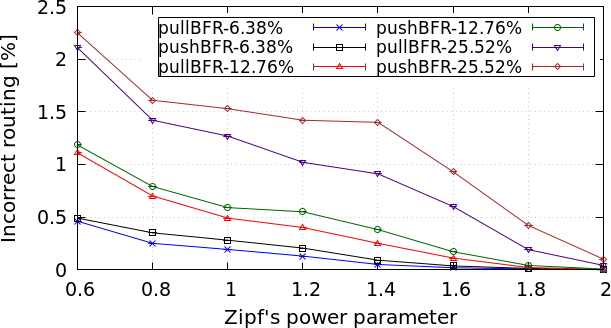}
 \caption{Perormance for different values of $\alpha$ in terms of the impact of false positive reports on routing.}
 \label{fig::fpImpact}
 \vspace{-1em}
\end{figure}

We analyze the impact of false positive reports on the performance of pull-based and push-based BFR. Fig. \ref{fig::fpImpact} compares these protocols in terms of the impact of false positive reports on routing. Using (\ref{MvaK}), we conducted experiments with $n=1000$ and three different rates for $p$ from set $F=\{6.38\%,12.76\%,25.52\%\}$ to observe the impact of false positive reports on the operation of our considered routing protocols. Fig. \ref{fig::fpImpact} shows the percentage of Interest messages that have reached wrong servers due to false positive reports from the BFs of CA messages at routers and clients. From Fig. \ref{fig::fpImpact}, we understand that the higher the value of $p$ is, the higher the percentage of incorrect routings is, for both pull-based and push-based BFR. The reason is that when we increase the value of $p$ for BFs, the probability that a false positive error occurs in practice is higher. Fig. \ref{fig::fpImpact} shows that the highest percentage of incorrect routing corresponds to $p=25.52\%$. However, even in this case, only $2.25\%$ of Interest messages have been routed towards wrong server(s). In practice, one will not use $p=25.52\%$ because it results in high risk of false positive reports. Fig. \ref{fig::fpImpact} makes clear that false positive reports have less impact on the operation of pull-based BFR compared to push-based BFR. This is because push-based BFR stores CA messages for the entire content universe. Hence, push-based BFR stores more CA messages compared to pull-based BFR, thus more number of BFs have to be checked. Further, from Fig. \ref{fig::fpImpact}, we observe that with higher $\alpha$ values, false positive reports have a smaller impact on the performance of both pull-based and push-based BFR. This is due to the fact that if the value of $\alpha$ is higher, smaller number of names are popular and, therefore, smaller number of names are checked against BFs, which results in less number of false positive reports. 
\section{Related Works}
\label{sec::rw}
In the recent years, many routing protocols have been proposed for NDN \cite{ospfn,nlsr,bfr,cobra,role,securenlsr}. Wang et al. propose OSPFN \cite{ospfn}, which is an extension to Open Shortest Path First (OSPF), as a routing protocol for NDN. OSPFN makes use of OSPF's Opaque LSAs \cite{opaqueospf} for advertising name prefixes in the routing messages. OSPFN considers the best next hop for each name prefix. However, it allows to consider alternative next hops as well. OSPFN has the following shortcomings: 1) it requires IP addresses to identify routers, 2) it requires to use ``GRE'' tunnels, and 3) it is a single-path routing protocol. 
 In \cite{nlsr}, Mahmudul Hoque et al. propose NLSR as a link state routing protocol for NDN, which uses LSA messages to exchange information about the available name prefixes as well as the topology of the network. NLSR proposes a hierarchical trust model to verify the LSA messages in a single domain. Lehman et al.  provide a description of NLSR features and its current design in \cite{securenlsr}. Push-based BFR \cite{bfr} is an intra-domain routing protocol for NDN, which operates based on BF-based content advertisements. 
 COBRA \cite{cobra} is a routing protocol, which has two operation phases namely learning phase and BF-based routing. In the learning phase, COBRA floods all the Interests. Rightafter, when Data packets are received, COBRA updates the route traces in Stable Bloom Filters (SBFs). If route traces for an Interest's name prefix is stored in SBFs, COBRA avoids flooding the Interest and routes the Interest according to the route trace information. The work in \cite{iscc} provides a comparative performance analysis of BFR and COBRA, where the advantages of BFR become more clear. 
\section{Conclusions}
\label{concl}
In this paper, we proposed pull-based BFR as a new routing protocol for NDN. Pull-based BFR has a number of advantages compared to push-based BFR: 1) significantly less communication overhead for propagating content advertisements, 2) BF-based aggregation mechanism for CAR and CA messages, 3) better average round-trip delay when $\alpha$ is in the $[1.2, 2]$ interval, 4) less storage space requirements for clients and routers to store content advertisements, and 5) more robustness to false positive reports from BFs. Similarly to push-based BFR, pull-based BFR is fully distributed, topology agnostic, content oriented, and does not need any IP-based routing protocol as a fall-back or primary routing mechanism.
For future work, we aim to further study scalability issues of pull-based and push-based BFR. 






\bibliographystyle{IEEEtran}
\bibliography{IEEEabrv,/home/ali/Documents/my_papers/ccnc/myrefs}
%
%
%

\end{document}